\documentclass[final,1p,times]{elsarticle}

\usepackage{graphicx}
\usepackage{amssymb}
\usepackage{amsthm}
\usepackage{lineno}
\usepackage{color}
\usepackage{soul}

\journal{Nuclear Physics A}

\begin{document}
\def\Journal#1#2#3#4{{#1} {\bf #2}, #3 (#4)}

\def\NCA{Nuovo Cimento}
\def\NIM{Nucl. Instr. Meth.}
\def\NIMA{{Nucl. Instr. Meth.} A}
\def\NPB{{Nucl. Phys.} B}
\def\NPA{{Nucl. Phys.} A}
\def\PLB{{Phys. Lett.}  B}
\def\PRL{Phys. Rev. Lett.}
\def\PRC{{Phys. Rev.} C}
\def\PRD{{Phys. Rev.} D}
\def\ZPC{{Z. Phys.} C}
\def\JPG{{J. Phys.} G}
\def\CPC{Comput. Phys. Commun.}
\def\EPJ{{Eur. Phys. J.} C}
\def\PR{Phys. Rept.}

\begin{frontmatter}

\title{Heavy Flavor, Quarkonia,
and Electroweak Probes at Quark Matter 2012}

\author[auth1]{Charles Gale}
\author[auth2]{Lijuan Ruan}
\address[auth1]{Department of Physics, McGill University, 3600 University Street, Montreal, QC, Canada H3A 2T8}
\address[auth2]{Physics Department, Brookhaven National laboratory, Upton NY, USA 11973}


\begin{abstract}
   We summarize and discuss some of the experimental and theoretical results on heavy flavor, quarkonia, and electro-weak probes presented at {\it Quark Matter 2012}.
\end{abstract}

\end{frontmatter} 


\section{Introduction}
This edition of the {\it Quark Matter} series of conferences has witnessed an impressive number of new theoretical and experimental results, which is a testament to the activity in this vibrant field. We discuss in turn some of the new results that were presented at the meeting.

\section{Theory results }
Starting with calculations of the electromagnetic emissivity of the quark-gluon plasma, photon emission rates have been known at leading order (LO)  in $\alpha_s$, the strong fine structure constant, for little over a decade \cite{Arnold:2001ms}. Given that RHIC phenomenology seems to favor values $0.2 \le \alpha_s \le 0.3$, one may wonder about the size of next-to-leading-order (NLO) contributions. These have recently been calculated and a report on the results has been given at this meeting \cite{DTeaney}.  The NLO contribution is parametrically of order $g^3 \ln (1/g) + g^3$, where $g$ is the strong interaction coupling constant.
Interestingly, the effect of small angle radiation gets enhanced at NLO, while the contributions at larger angles - of the same order in $g$ - get suppressed at NLO, owing in part to an interplay between enhanced scattering rates and a modification of the phase space.
\begin{figure}[tbhp]
\begin{center}
\includegraphics[trim=0cm -.05cm 0cm 0cm, clip, width=0.331\textwidth]{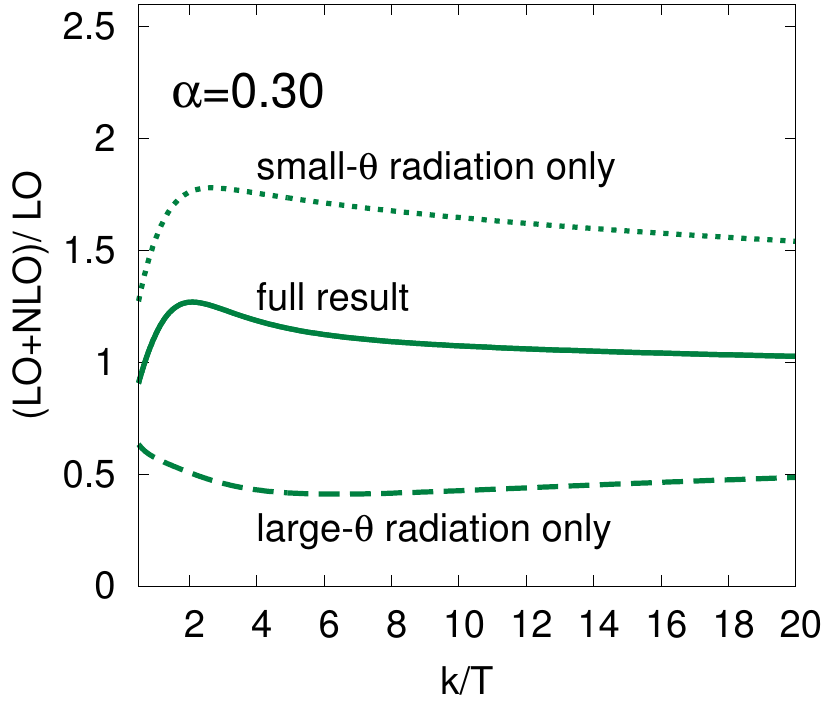}\hspace*{1cm}
\includegraphics[trim=0cm -.8cm 0cm 0cm, clip, width=0.382\textwidth]{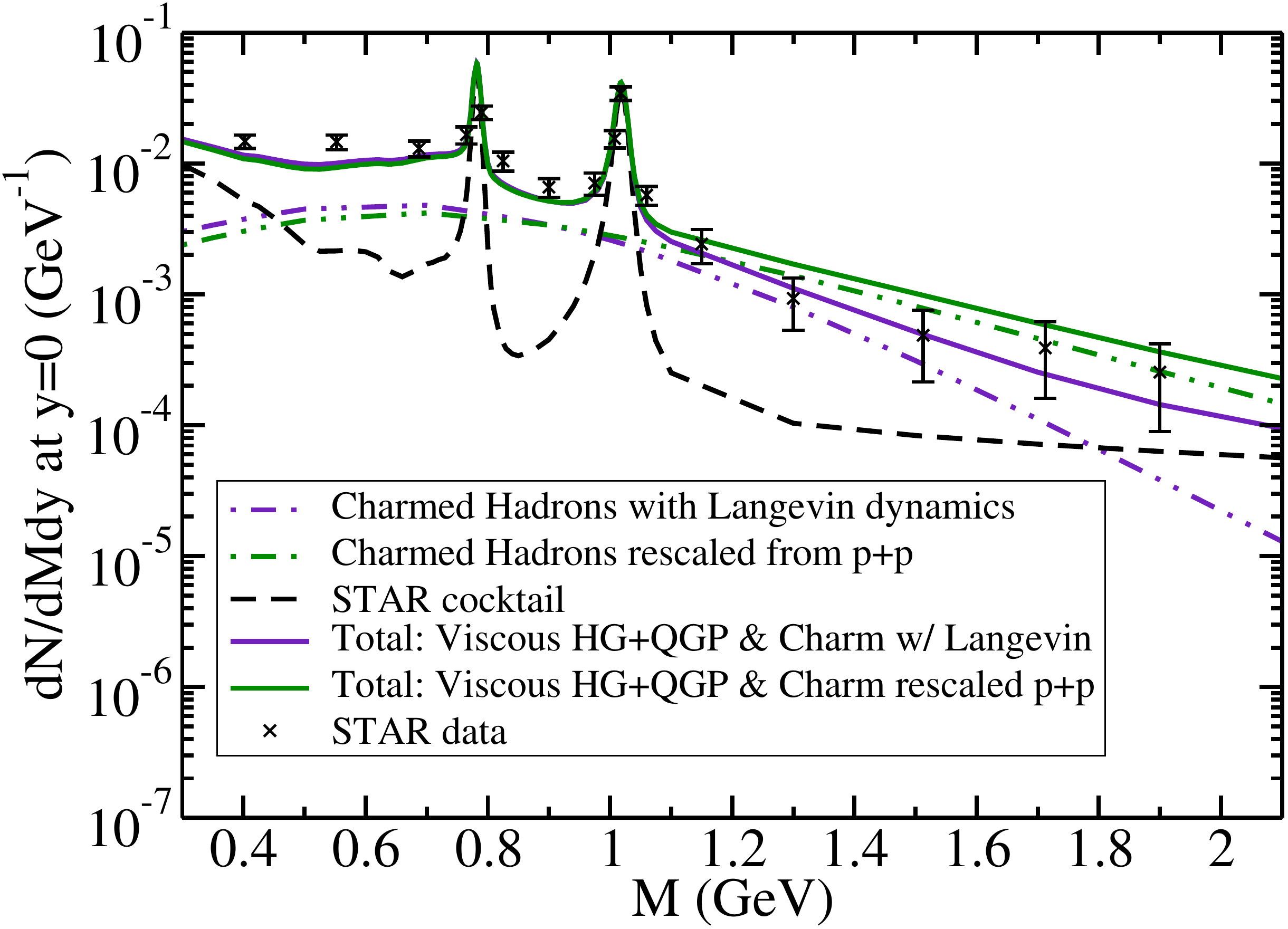}
\vspace*{-0.5cm}
\end{center}
\caption{(Color Online)
(Left panel) The ratio (LO+NLO)/LO for the photon production rate, with parameters appropriate for RHIC phenomenology ($\alpha_s = 0.3$, $N_f =3$). This plot is from Ref. \cite{DTeaney}. (Right panel) The dilepton invariant mass spectrum, obtained using a 3+1D viscous hydrodynamic model. The full lines show the effect of charm quark energy loss (top/bottom: without/with energy loss). This plot is from Ref. \cite{Gojko}.}
\label{Phot_NLO}
\end{figure}
The net effect, shown in the left panel of Figure \ref{Phot_NLO}, is a modest  ``K-factor'', which deviates from unity by  at most 20-30\%  at  $k/T \sim 2$. Bearing in mind the caveats inherent to a perturbative analysis, this NLO study implies that the phenomenology that relies on those photon rates need not be redone. The techniques developed there, however, will be of use for other finite temperature applications. Other communications on electromagnetic  observables included a proposal to observe photons originating from QCD jet conversion by first triggering on a reconstructed jet, and then searching for the back-scattered photon correlated with this hard parton \cite{Fries}. This technique should avoid contamination from the direct photon sources without an underlying hard process, and suppress fragmentation photons which populate low values of $z = E^\gamma/E^{\rm jet}$. Recent dilepton calculations were also shown which highlighted the effects of a finite shear viscosity coefficient on lepton-pair observables \cite{Gojko}. The dilepton yields were shown to be somewhat insensitive to the viscosity  corrections, but the dilepton elliptic flow - as quantified by $v_2 (p_T)$ - shows greater sensitivity. The absolute value of the predicted dilepton  $v_2$ does remain small, which is  consistent with the different   hydrodynamic calculations of photon elliptic flow \cite{Rupa:09}. These, however, are at odds with current  measurements by the PHENIX collaboration. The theoretical results for dilepton yields were in accord  with recent measurements by the STAR collaboration (to be shown and discussed further below), and relied on vector spectral densities that were also found to be consistent with NA60 data \cite{Ruppert:2007cr}. 
 The data there appear to be consistent with a scenario where the charm quarks  loose energy during the viscous evolution, see Figure \ref{Phot_NLO}, right panel. All channels of energy loss will need to be considered before a final assessment is made.
\begin{figure}[htbp]
\begin{center}
\includegraphics[width=0.39\textwidth]{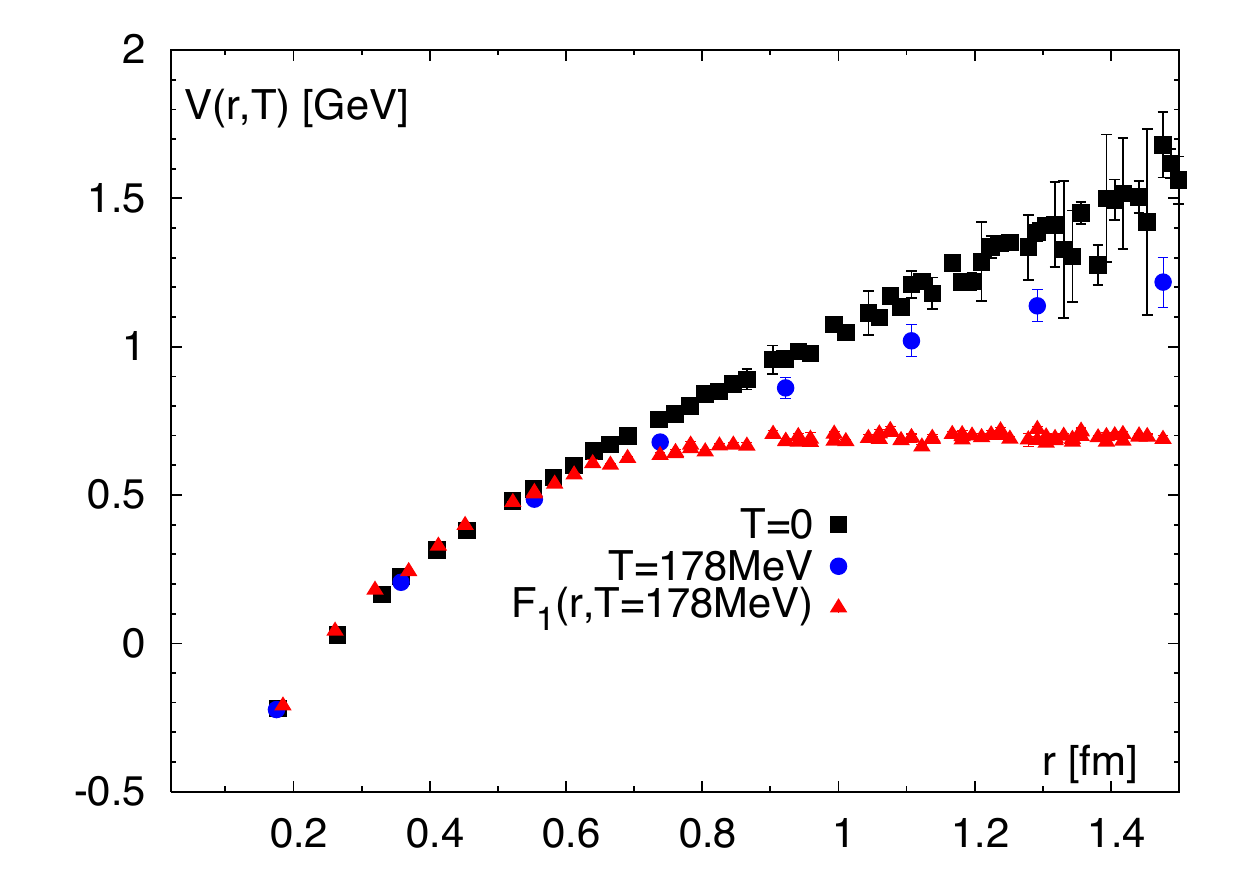}\hspace*{1.3cm}
\includegraphics[width=0.39\textwidth]{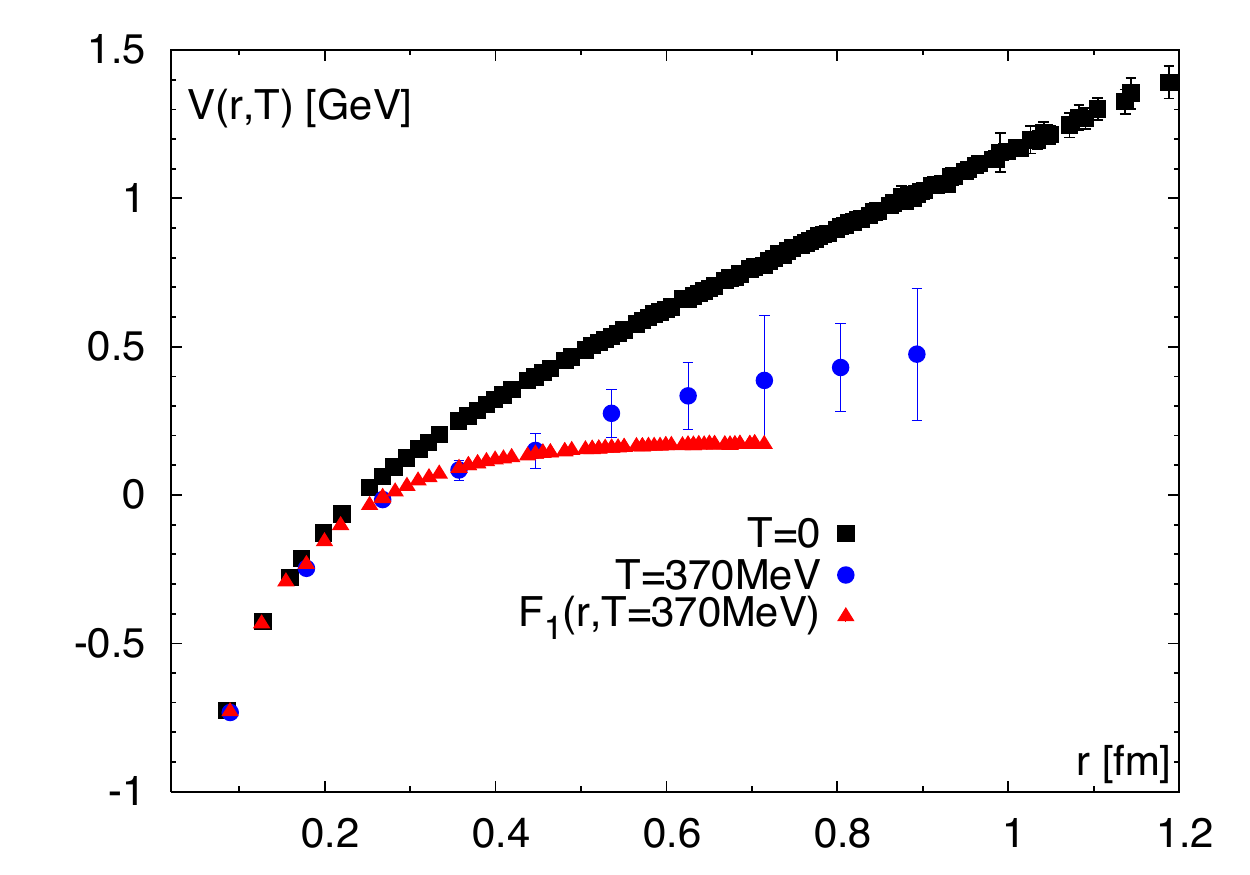}
\end{center}
\caption{(Color Online) The non-perturbative finite-temperature quark-antiquark potential, show together  with the $T=0$ potential, and with the singlet free-energy, $F_1$. (Left panel) $T = 178$ MeV, (Right panel) $T = 370$ MeV. This plot is from Ref. \cite{Bazavov}.}
\label{QQpot}
\end{figure}

In what concerns non-perturbative techniques, advances in lattice QCD calculations were reported at the meeting. One result highlights  progress in the study of quarkonium suppression as a signal of the quark-gluon plasma (QGP). It addresses some of the ambiguities in extracting a static quark potential at finite temperatures, by identifying the static $Q\bar{Q}$ energy using a spectral decomposition of the temporal Wilson loop \cite{Bazavov}. That work shows, on Figure \ref{QQpot}, that the singlet free energy and the actual potential are only similar at high temperatures, and at that temperatures commensurate with that of the crossover transition the potential deviates only slightly from it zero temperature behavior. This step forward should impact some of the phenomenological studies presented at the meeting \cite{Rapp} (some of these results will be shown in sections to follow).
\begin{figure}[htbp]
\begin{center}
\includegraphics[width=0.4\textwidth]{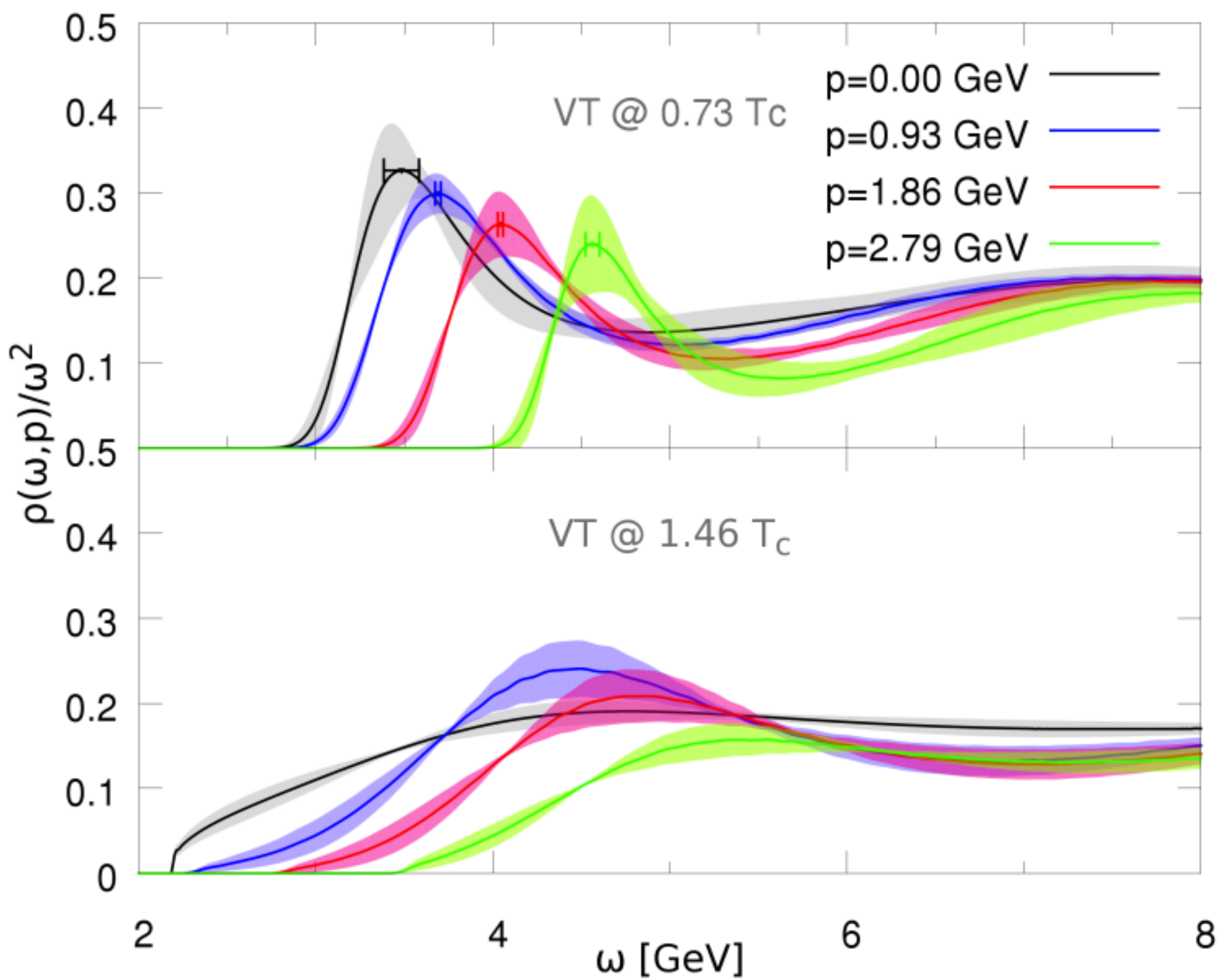}
\includegraphics[width=0.4\textwidth]{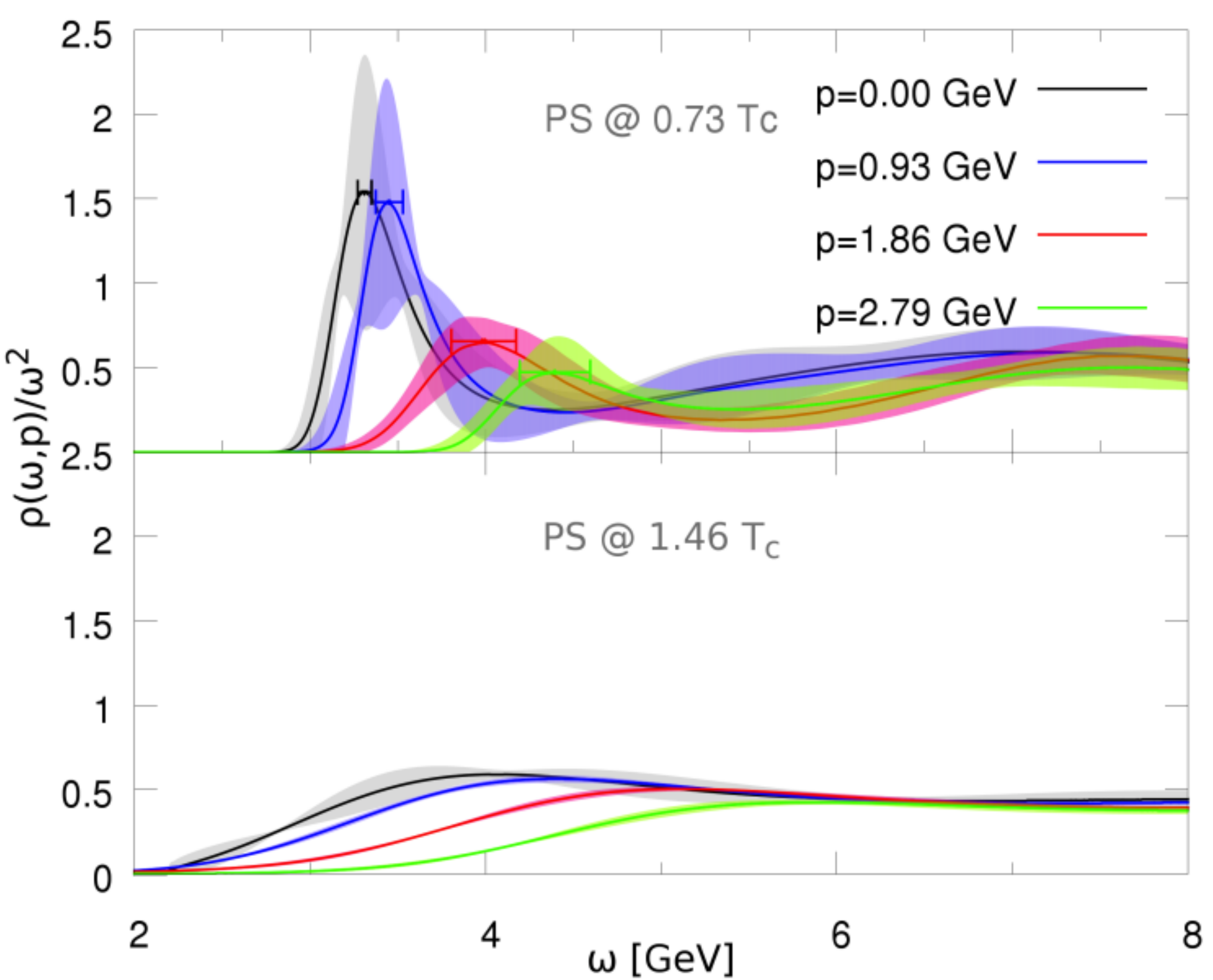}
\end{center}
\caption{(Color Online) The charmonium spectral densities, plotted for different momenta against the energy. The vector (left panel) and pseudo-scalar (right panel) channels are shown  at different values of the temperature. This plot is from Ref. \cite{Ding}.}
\label{Spectral}
\end{figure}
Still in the context of the in-medium modifications of quarkonium, another lattice communication reported on developments  in the determination of the {\it dynamical} properties of the charmonium spectral density \cite{Ding}.   By extracting the spectral profiles through the Maximum Entropy Method, one can obtain the transverse component of the vector channel, and the pseudo scalar channel, both of which  are shown in Figure \ref{Spectral}. A clear momentum dependence is observed in both vector and pseudo-scalar spectral densities at the temperatures considered here, and dissolution is observed at $T = 1.46\, T_c$ in this quenched approximation . Clearly, calculations at other temperatures are needed to complete the description of the fate of charmonium moving through the heat bath, and to make contact with results obtained from using effective field theory or AdS/CFT techniques. Finally, results of calculations of $J/\psi$ suppression were shown, where the short-distance $Q\bar{Q}$ production cross-sections were evaluated using non-relativistic QCD (NRQCD), and the longer distance in-medium effects were incorporated through formation and dissociation times for the quarkonium states. Thermal effects, neglected in the bound state wave function, appear necessary to interpret the current measurements, especially at the  LHC \cite{Sharma}.

\begin{figure}[htbp]
\begin{center}
\includegraphics[width=0.35\textwidth]{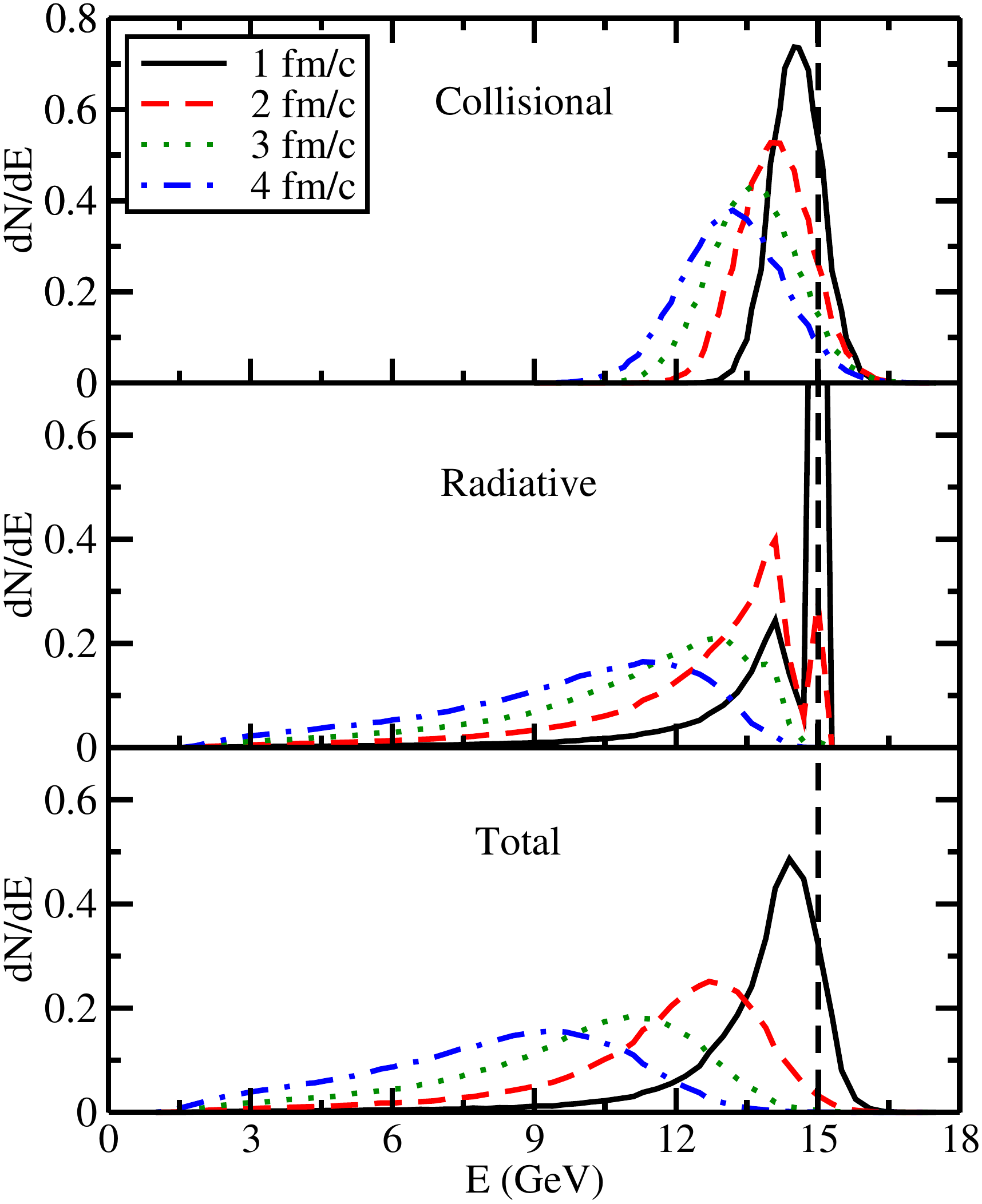}\hspace*{1cm}
\includegraphics[width=0.4\textwidth]{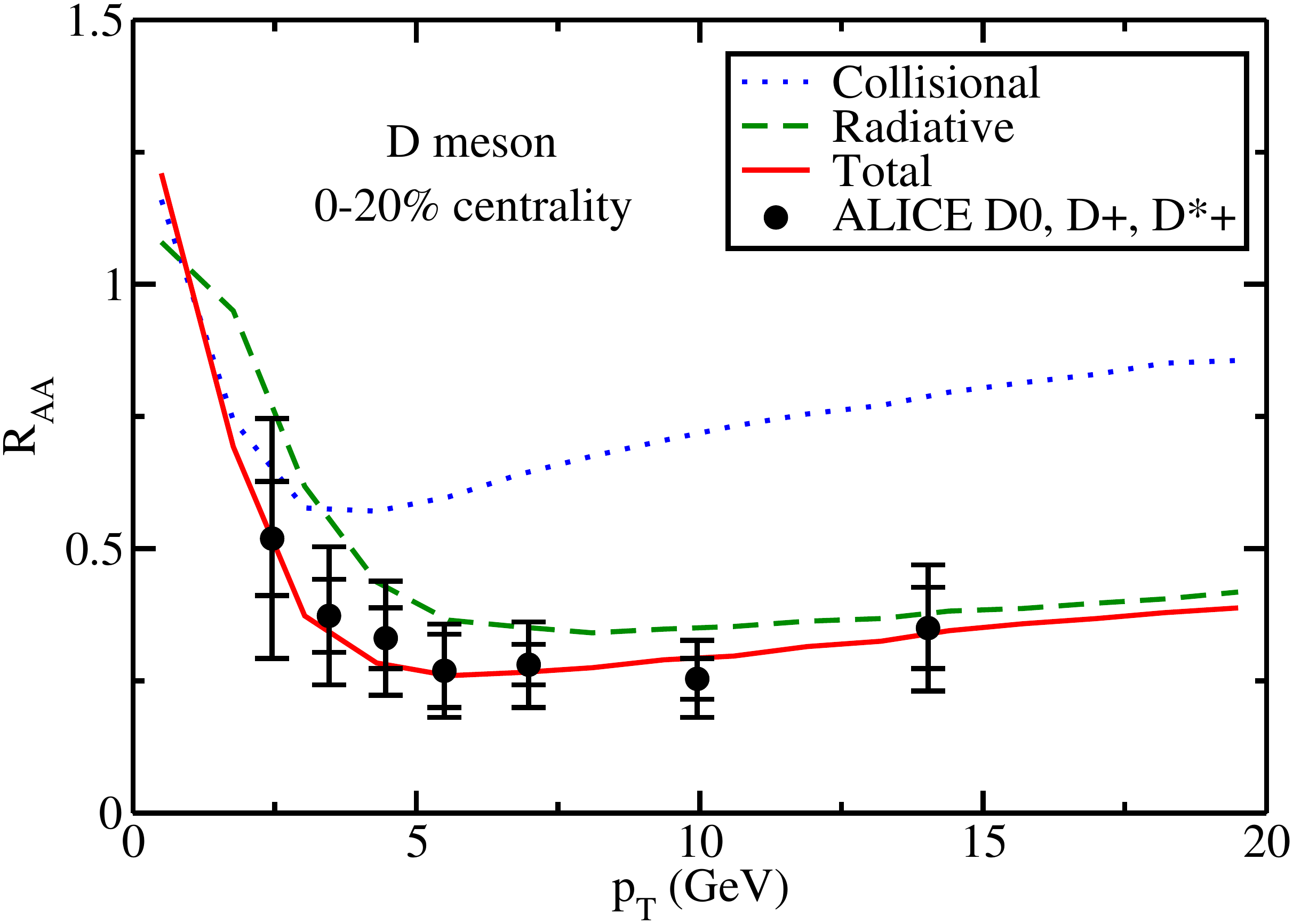}
\end{center}
\caption{(Color Online) (Left panel) The time-dependent energy profile of a heavy quark, originally with $E = 15$ GeV. Effects of collisional, radiative, and net energy loss are shown separately. (Right panel) The individual and combined effect of the collisional and radiative energy loss on $D$ meson $R_{AA}$, plotted together with ALICE results. Both plots are from \cite{Cao}.}
\label{Eloss}
\end{figure}
The properties of heavy quarks in the hadronic medium - either cold or hot - were the subject of several talks reporting new results. These included an attempt to calculated cold nuclear matter effects from first principles, and to estimate gluon saturation effects in charmonium production in d+A and A+A collisions \cite{Tuchin}.
The results of a study on the evolution of heavy quarks in an evolving viscous medium were also shown. This evolution relied on Langevin dynamics modified to also include radiative energy loss \cite{Cao} (usually neglected), calculated in the higher-twist formalism \cite{HT}. The time-dependence of heavy quark energy profile is especially relevant to lepton pair  studies, as charm quark decays - for example - form an irreducible dilepton background, as discussed further in subsequent sections. The left panel of  Figure \ref{Eloss} shows the evolution in time of the energy spectrum of a 15 GeV  heavy quark, in a static QCD ``brick'' with $T = 300$ MeV. Clearly, the initial delta function is quickly broadened and shifted in energy, the evolution is inherently non-linear, and the radiative losses are important. This point is driven home in the right panel, where both effects are needed to achieve a satisfactory understanding of LHC data of $D$-meson $R_{AA}$.

In summary, some fundamental theoretical results were shown at this meeting, which reflected advances in our understanding of fundamental processes, as well as in the detailed modelling of relativistic nuclear collisions.

\section{Experimental results}
\subsection{Heavy flavor}
There were many new results shown for heavy flavor measurements,
including those for $D$ and $B$ mesons, and their decay leptons.  It is found that at
RHIC, the charm cross section follows the number of binary collisions scaling
from $p+p$, to d+Au, to Au+Au collisions at $\sqrt{s_{_{NN}}} = 200$
GeV~\cite{wei:12}. 
Nuclear modification factors $R_{AA}$ of $D^{0}$ at 200 GeV Au+Au
collisions show a strong $p_T$ dependence~\cite{wei:12}. They
increase as a function of $p_T$, reach a maximum value at
$p_T\sim1$ GeV/$c$ and then decrease. At LHC, the $R_{AA}$ of charmed
mesons are measured for $1\!<\!p_T\!<30$ GeV/$c$ in Pb+Pb
collisions at 2.76 TeV~\cite{valle:12,Grelli:12}. For $p_T\!>4$ GeV/$c$, the observed suppression factors
of charmed mesons are similar to those of pions at RHIC
and LHC. Results from non-prompt $J/\psi$ measurements indicate
that $R_{AA}(B\rightarrow J/\psi)$ is larger than $R_{AA}(D)$ in
central Pb+Pb collisions~\cite{Mironov:12}. More precise measurements together with
advanced theoretical calculations should  enable us to understand
the color and/or flavor dependence of energy loss in greater depth. In
addition, the $D$ meson shows a significant positive elliptic flow
($v_2$) for $2\!<\!p_T\!<\!10$ GeV/$c$~\cite{Caffarri:12}. These measurements indicate
that heavy flavor quarks interact with the medium strongly, which
leads to a similar magnitude of suppression as for pions and generates
a significant flow.
\subsection{Quarkonia: interplay of color screening and recombination}
Heavy quarkonia are thought to be an excellent probe to study
color screening features in a hot, dense medium. Different
quarkonium states are predicted to dissolve at different
temperatures, owing to their different binding energies~\cite{rhicIIQuarkonia}. Precise
measurements of $p_T$ distributions of quarkonia at different
centralities, collision systems and energies should serve as a
thermometer of the QGP. However, in addition to color screening
features, there are also many effects playing an important role
such as gluon dissociation, heavy quark recombination, hot-wind
dissociation, formation time, and cold nuclear matter effects~\cite{rhicIIQuarkonia}. 

Collisions from asymmetric systems could provide additional information on cold nuclear matter effects. It is found that in 200 GeV Cu+Au collisions, $R_{AA}$ of $J/\psi$ in the
Au-going direction is larger than that in the Cu-going
direction. A model taking into account cold nuclear matter
effects including parton distribution functions and the $J/\psi$
break-up cross section qualitatively describes the feature, but
under-predicts the magnitude of the suppression~\cite{Rosati:12}. The additional
suppression suggests hot, dense medium effects.

\begin{figure}[htbp]
\begin{center}
\includegraphics[width=0.35\textwidth]{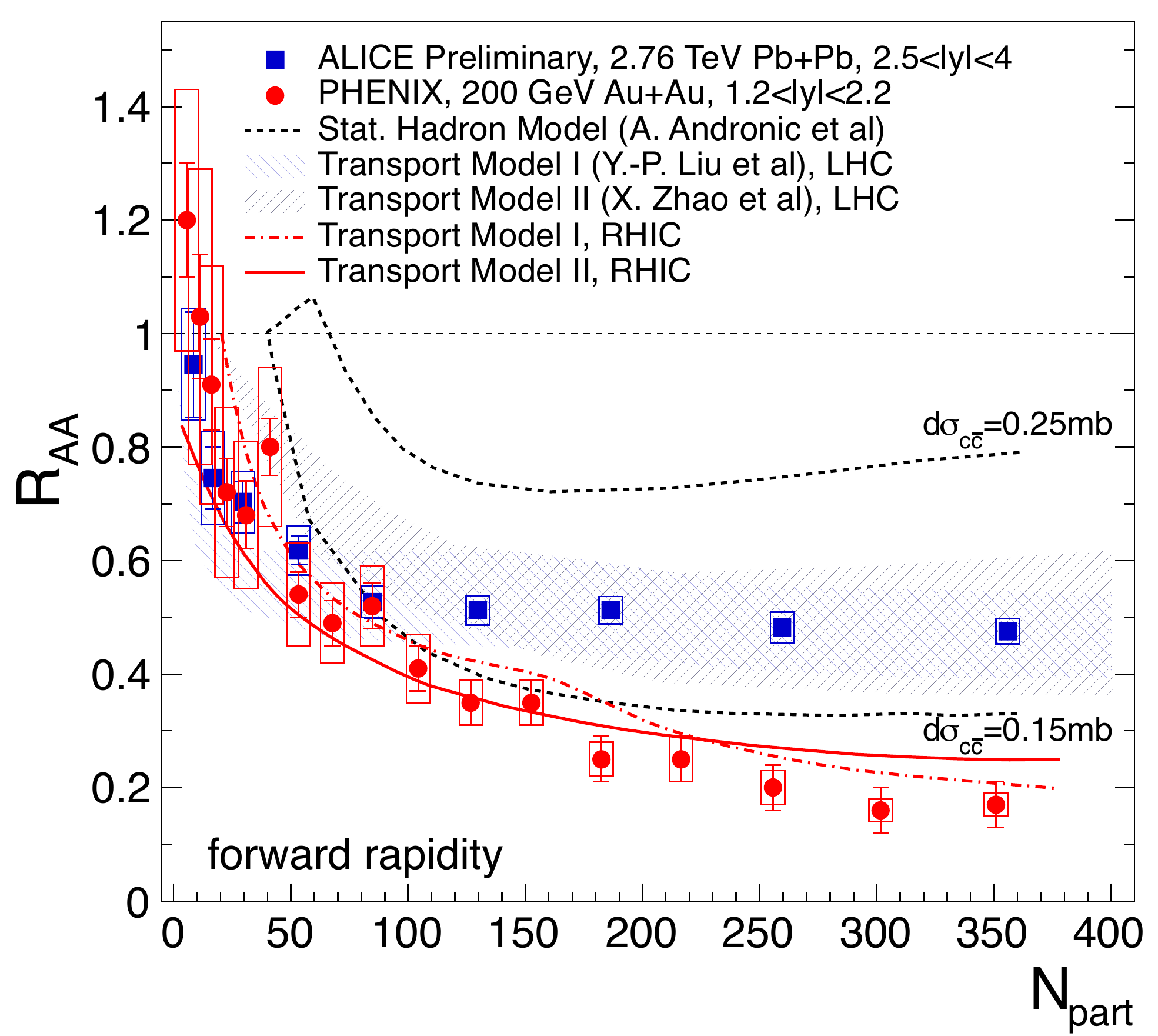}
\includegraphics[width=0.35\textwidth]{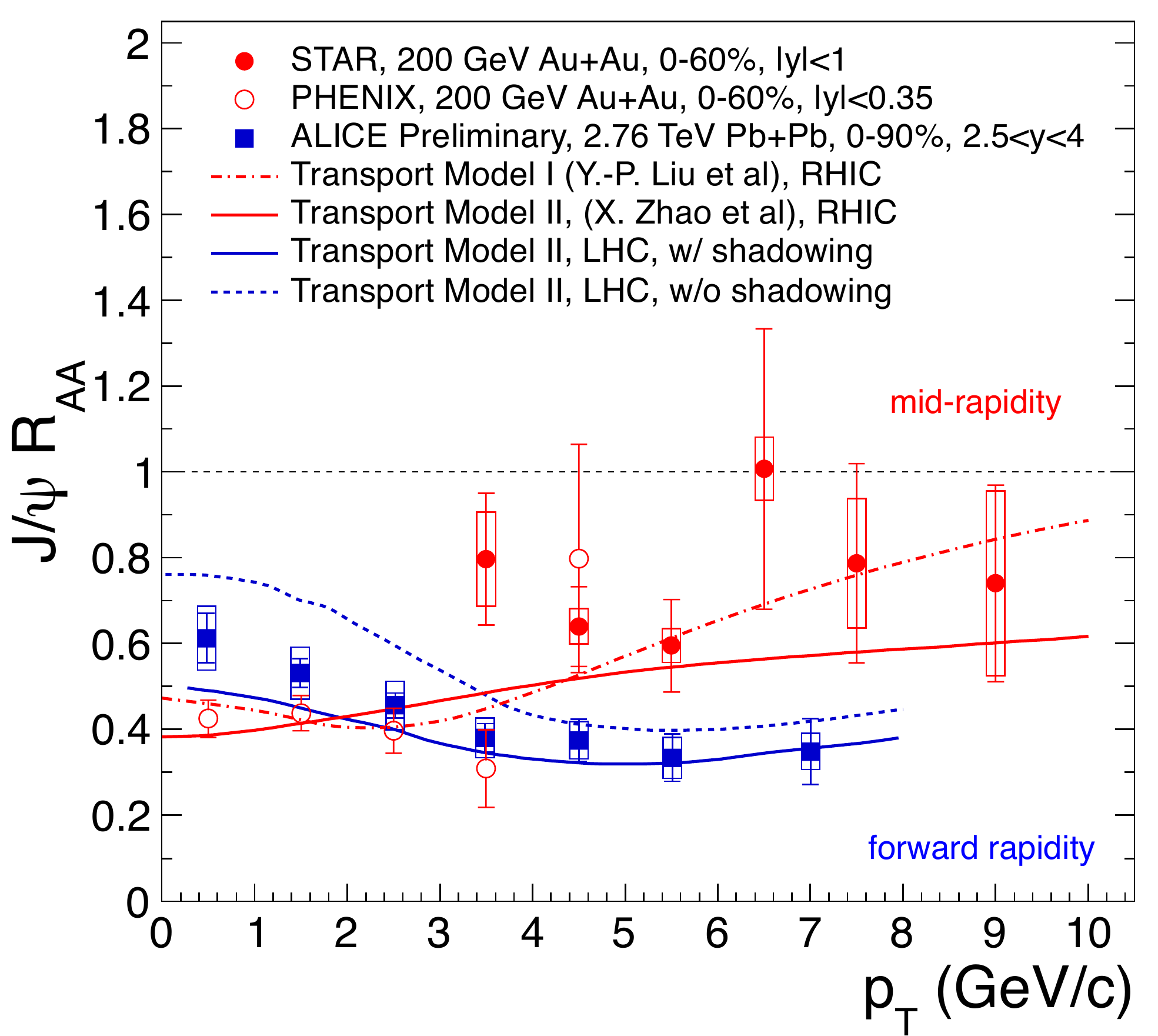}
\includegraphics[width=0.35\textwidth]{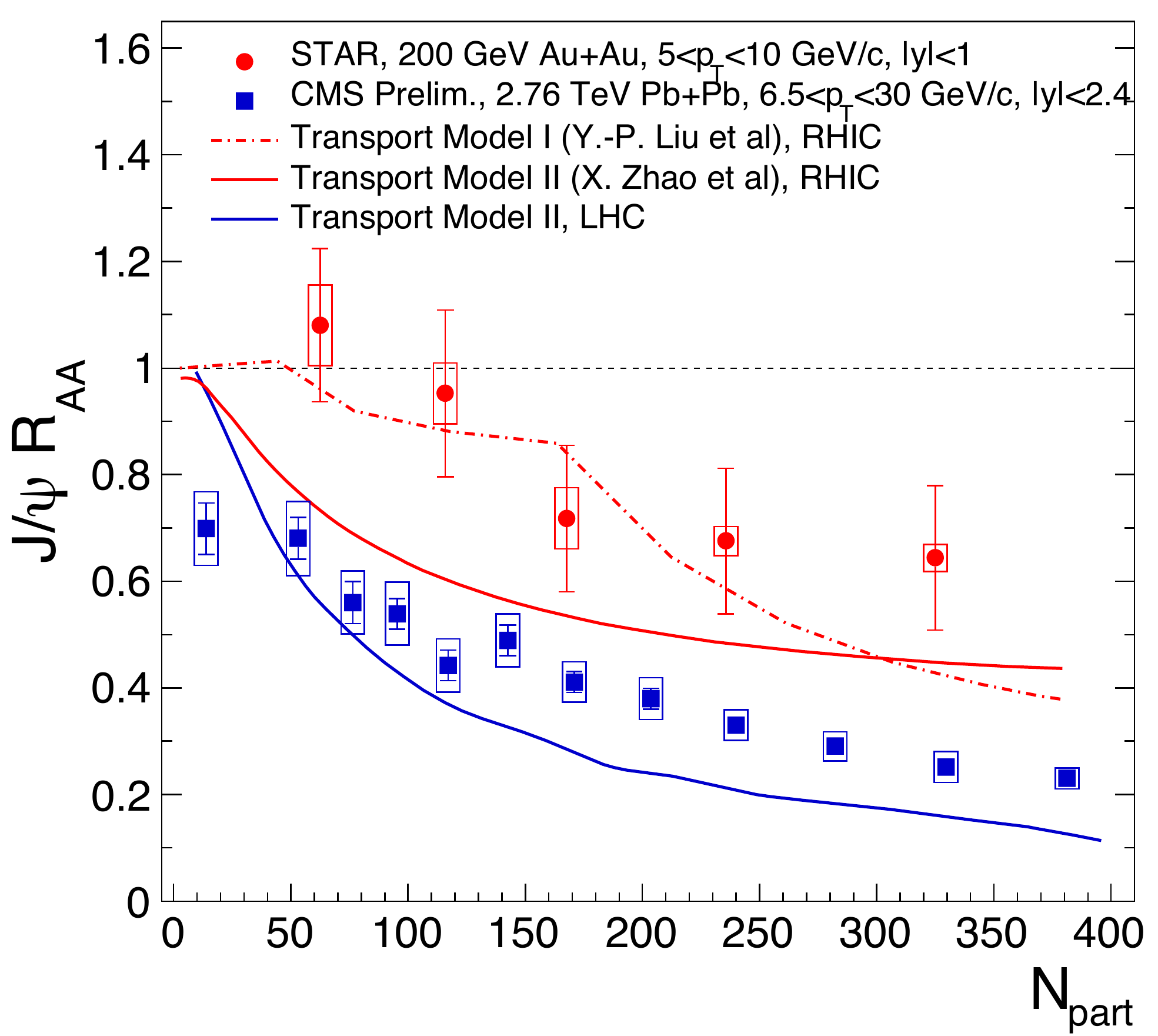}
\includegraphics[width=0.35\textwidth]{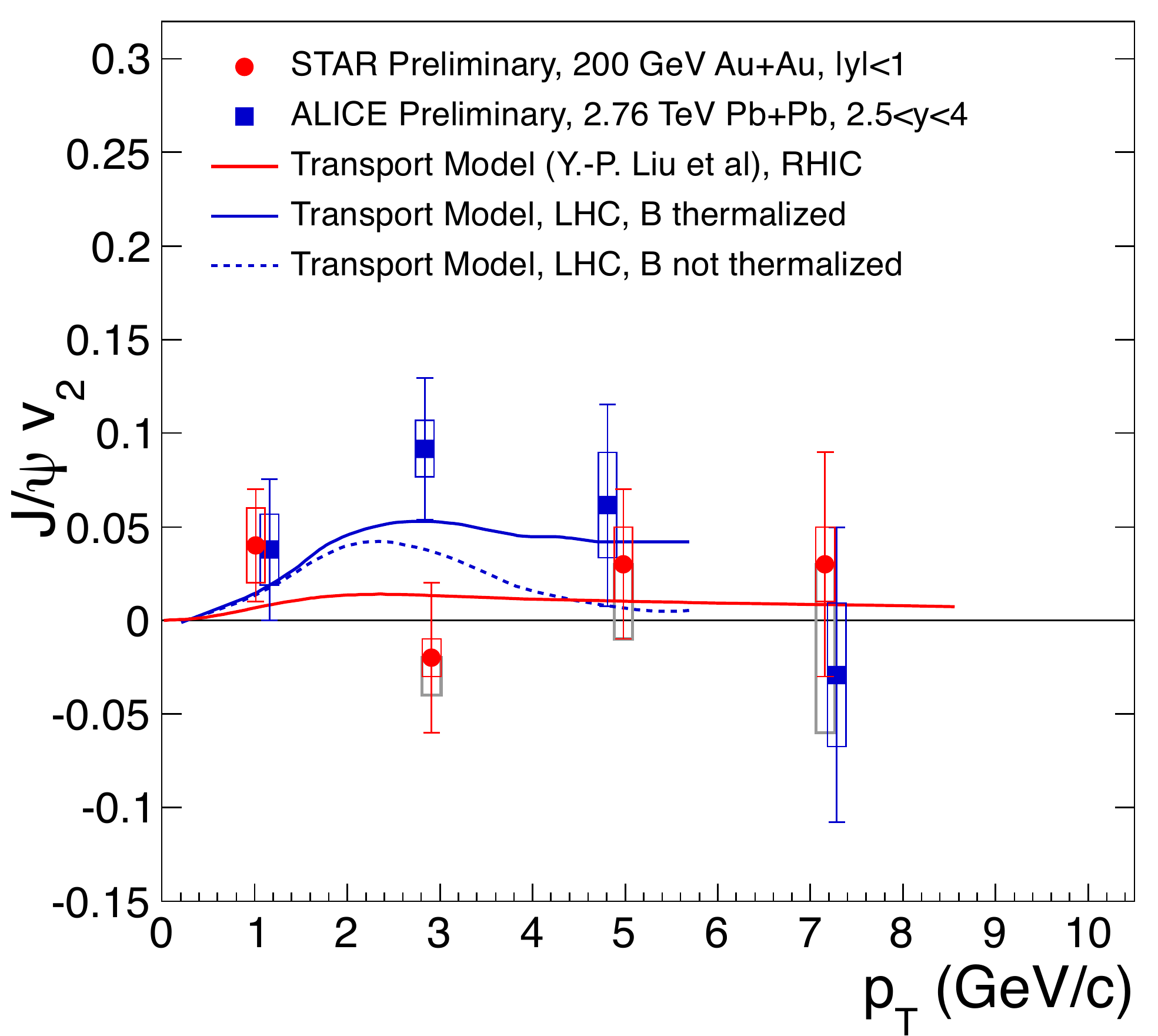}
\end{center}
\caption{Color Online (upper-left panel): $J/\psi$ $R_{AA}$ as a
function of centrality in forward rapidity at RHIC and LHC.
(upper-right panel): $J/\psi$ $R_{AA}$ as a function of $p_T$.
(bottom-left panel): High $p_T$ $J/\psi$ $R_{AA}$ as a function of
centrality. (bottom-right panel): $J/\psi$ $v_2$ as a function of
$p_T$. The normalization uncertainties are not shown. Model comparisons are shown as curves and bands.} \label{fig:1}
\end{figure}

Figure~\ref{fig:1} shows systematic comparisons of $J/\psi$
suppression patterns at RHIC and LHC. In central collisions,
$J/\psi$ is less suppressed in Pb+Pb at 2.76 TeV  for
$2.5\!<|y|\!<4$ than in Au+Au collisions at 200 GeV for
$1.2\!<|y|\!<2.2$~\cite{phenixjpsi, ALICEjpsi}. A similar feature is observed for mid-rapidity
comparisons between LHC and RHIC energies. Comparison to model calculations indicates that
recombination plays a more significant role at LHC \cite{Rapp}. Furthermore,
the $p_T$ dependence of $J/\psi$ suppression~\cite{ALICEjpsi, phenixjpsi:07, starjpsi:12}, the flow pattern of
$J/\psi$ as a function of $p_T$~\cite{Yang:12, Trzeciak:12}, and the high $p_T$ $J/\psi$
suppression pattern as a function of centrality~\cite{starjpsi:12,Moon:12}, reported at LHC
and RHIC, can be described consistently by model calculations
incorporating color screening and recombination features. We note
that substantial suppression observed at high $p_T$ points to the
color screening feature since the recombination and cold nuclear
matter effects are negligible there.

A cleaner probe to study the color screening feature is $\Upsilon$
since the $b\bar{b}$ recombination is small. Measurements at
CMS show that $R_{AA}(1S)$ is 0.4-0.5 while $R_{AA}(2S)$ is 0.1 in central Pb+Pb collisions at 2.76 TeV~\cite{cmsupsilon:12}. At 200 GeV,
$R_{AA}(1S+2S+3S)$ is about 0.4 in central Au+Au collisions~\cite{Trzeciak:12}. Considering
feed-down contributions [$\sim$50\% for $\Upsilon(1S)$],
measurements at RHIC and LHC indicate that $\Upsilon(3S)$ is
completely melted and that $\Upsilon(2S)$ is strongly suppressed
in central A+A collisions. In the future, different $\Upsilon$ states should be measured at RHIC with the Muon Telescope Detector (MTD) upgrade at STAR and sPHENIX upgrade at PHENIX~\cite{mtd, sphenix}. In light of RHIC
and LHC precise quarkonium measurements, we are in the era to study
color screening features of hot, dense medium.

\subsection{Electro-weak probes}
W and Z bosons and high energy direct or isolated photons are ideal
probes to study the initial nuclear wave function effect since
they are produced at initial impact and go through the medium with
minimum interactions. At 2.76 TeV, the W and Z boson production is
found to follow number of binary collisions scaling from $p+p$ to Pb+Pb
collisions~\cite{cmswzgamma, atlaswzgamma}. The isolated photon production also follows binary
scaling for $20\!<\!p_T\!<\!200$ GeV/$c$ at mid-rapidity~\cite{cmswzgamma, atlaswzgamma}. At 200 GeV,
$R_{AA}$ for direct photons is consistent with unity within
uncertainties at $5\!<\!p_T\!<\!18$ GeV/$c$ in Au+Au
collisions at mid-rapidity~\cite{phenixgamma}. These measurements all suggest that
initial-state effects are small for
electro-weak probes.

\subsubsection{Thermal photon spectra and $v_2$}
In addition to high energy photon measurements, lower energy
photons are used to study thermal radiation. For
$1\!<\!p_{T}\!<\!4$ GeV/$c$, PHENIX measured direct photon yields
from di-electron measurements and found an excess in 0-20\% Au+Au over $p+p$ at $\sqrt{s_{_{NN}}} = 200$ GeV, which is exponential in $p_T$ with slope parameters
221 MeV~\cite{thermalphoton}. At LHC, an excess of direct photon yield in 0-40\% Pb+Pb
collisions above $p+p$ was reported at $\sqrt{s_{_{NN}}} = $ 2.76
TeV, exponential in $p_T$ with slope parameters 304 MeV~\cite{ALICEphoton}. If
indeed the excess is from the QGP phase, the measurements at RHIC and
LHC would indicate that the initial temperature for the QGP
evolution is as high as 300-600 MeV~\cite{thermalphoton}.

On the other hand, $v_2$ has been measured for direct photons and
found to be substantial in the range $1\!<\!p_{T}\!<\!4$ GeV/$c$
in central 0-20\% Au+Au collisions at $\sqrt{s_{_{NN}}} = 200$
GeV~\cite{photonv2}. Model calculations~\cite{Rupa:09} for QGP
thermal photons in this kinematic region significantly
under-predict the observed $v_2$, though if a significant
contribution from the hadronic sources at later stages is added,
the excess of the spectra and the observed $v_2$ at
$1\!<\!p_{T}\!<\!4$ GeV/$c$ are described reasonably
well~\cite{rapp:11}. It has been proposed that dilepton $v_2$
measurements
as a function of
$p_T$ in different mass regions would enable us to probe the
properties of medium from a hadron-gas-dominated to a QGP-dominated scenario~\cite{Gale:07}.

\subsubsection{Dilepton spectra and $v_2$}
At QM2012, STAR reported the di-electron $v_2$
measurements from 200 GeV Au+Au collisions. Within uncertainties, the data are compatible  with simulations containing contributions from known hadronic sources without QGP or hadron
gas thermal radiation~\cite{geurts:12,cui:12}. However, 
with more statistics, the di-electron $v_2$
measurement will provide additional sensitivity to study the thermal
radiation from the different phases.

\begin{figure}[htbp]
\begin{center}
\includegraphics[width=0.7\textwidth]{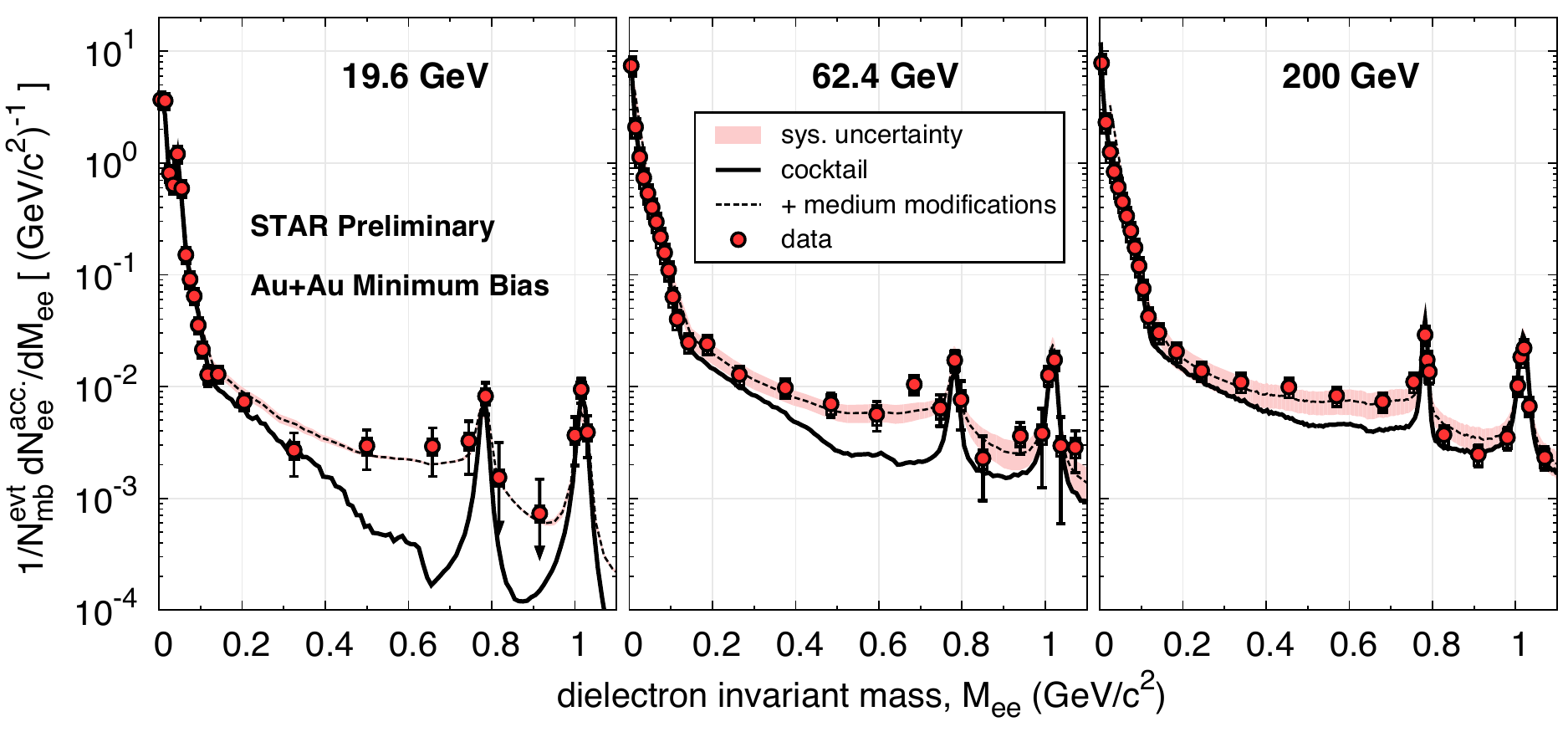}
\end{center}
\caption{(Color Online) Di-electron spectra in Au+Au collisions
from 19.6, 62.4 and 200 GeV from STAR. Comparisons to model
calculations with a broadened spectral function are also shown.}
\label{fig:2}
\end{figure}

We now come to measurements of dilepton spectra.
At RHIC, the PHENIX experiment observed a significant enhancement
in the $e^{+}e^{-}$ data above the expectation from hadronic
sources for $0.15\!<\!M_{ee}\!<\!0.75$ GeV/$c^{2}$ for $p_T\!<1$
GeV/$c$ in 200 GeV Au+Au collisions~\cite{lowmass}. Models~\cite{rapp:09,PSHD:12,USTC:12} that successfully
describe the SPS dilepton data consistently fail to describe the
PHENIX data. With the Time-of-Flight detector upgrade, STAR
reported the di-electron spectra in 200 GeV Au+Au collisions at
QM2011 and the low-mass excess was not as significant as what
PHENIX observed in 0-80\% and 0-10\% collisions~\cite{ppdilepton:12,Jie:11}. Further
comparisons point to the fact that the discrepancy between STAR
and PHENIX is in collisions in the 0-20\% centrality class only. 

At this conference, PHENIX reported the di-electron results from
the Hadron Blind Detector from 20-92\% Au+Au collisions at 200
GeV, which are consistent with those in the previous publication~\cite{phenixdielectron:12}.
STAR reported the di-electron spectra from 19.6, 39, and 62.4 GeV
Au+Au collisions~\cite{geurts:12,huang:12}. A broadened spectral function~\cite{rapp:09}, which describes
SPS dilepton data, consistently accounts for the STAR low mass
excess at 19.6, 62.4 and 200 GeV, as shown in Fig.~\ref{fig:2}. See also Ref. \cite{Gojko}.
Is there a connection to chiral symmetry restoration?
Chiral condensates,
which have a direct connection to spectral functions, strongly
depend on total baryon density and temperature.
To establish a link between dilepton measurements and chiral symmetry restoration,
we need more theoretical efforts and more precise measurements
over a broad beam energy range from FAIR, RHIC and LHC.
In the future, e-muon correlations from the MTD upgrade will measure
the correlated charm contribution~\cite{mtd, ruan:12}. This should enable us to obtain the QGP thermal radiation signal using the intermediate-mass dilepton measurements at RHIC.

\section{Summary}
This edition of the Quark Matter conferences has seen a stream of results unprecedented in their quality and in their quantity, being presented by the theory community, and by the experimental collaborations working at RHIC and at the LHC. Calculations have brought us significantly closer to an {\it ab-initio} modelling of relativistic heavy ion collisions. The precise charm and bottom
measurements will constrain heavy flavor dynamics in hot, dense
medium. Precise quarkonium measurements put us into the era of studying and characterizing color screening properties.
Thermal photon and dilepton measurements from a broad beam energy range enable us to study the fundamental properties of QGP, of chiral symmetry restoration, and will provide stringent tests of the dynamical evolution scenarios  of relativistic nuclear collisions.\\

\noindent {\bf Acknowledgment} The work of CG is supported in part by the Natural Sciences and Engineering Research Council of Canada, and the work of of LR is supported in part by the U. S. Department of Energy under Contract No. DE-AC02-98CH10886.


\end{document}